\documentstyle[preprint,eqsecnum,aps]{revtex}
\tighten
\begin{document}
\draft

\preprint{NCL93-TP13}
\title{Black hole formation from colliding bubbles}
\author{Ian G. Moss}
\address{Department of Physics, University of Newcastle upon Tyne,
NE1 7RU,United Kingdom}
\maketitle
\begin{abstract}
Some indication of conditions that are necessary for the formation
of black holes from the collision of bubbles during a supercooled
phase transition in the the early universe are explored. Two colliding
bubbles can never form a black hole. Three colliding bubbles can
refocus the energy in their walls to the extent that it becomes
infinite.
\end{abstract}
\pacs{98.80Cq, 04.60.+n, 97.60.Lf}
\date{}

\section{INTRODUCTION}

A small number of black holes being produced in the early stages of
universe would have important cosmological consequences, both from
their contribution to the density of the universe and to the high
energy cosmic ray background \cite{hawking0,carr,halzen}. For this to
have occured there would have to have been density perturbations with
large amplides because small amplitude perturbations do not grow in
size in the primordial plasma. The isotropy of the universe suggests to
us that the universe was fairly homogeneous on the length scales
corresponding to galaxies or larger objects, but comparatively little
is known about the homogeneity of the early universe on small scales,
for example scales much less than the horizon scale at nucleosynthesis
which encompassed a mass of around $3\times 10^{36}Kg$.

One situation in which a substantial amount of inhomogeneity may have
arisen could be at a first order phase transitions where a gauge
symmetry was broken. The possibility of black hole formation was raised
very early in the theory of these phase transitions
\cite{sato,hawking}.

First order phase1)       transitions are characterised by tunnelling from the
old to the new phase. For spontaneous symmetry breaking the phases are
distinguished by the values of a Higgs field. The tunnelling is modeled
by the instantaneous nucleation and expansion of a bubble of the
broken--symmetry phase \cite{coleman,guth}. In the supercooled case,
where the tunnelling is smaller than the expansion rate of the
universe, it is possible for the thermal energy of the old phase to be
surpassed by the energy of the vacuum, then the bubble walls accelerate
to speeds close to the speed of light and build up considerable amounts
of energy.

The collision of these bubbles and the possibility of forming black
holes was considered in reference \cite{hawking}. We found that when
two bubbles collide the bubble walls could continue through one another
unchanged or the energy could convert into the phase of the Higgs
field. In the latter case, most of the energy emerged from the
collision region as phase waves moving at the speed of light. We also
argued on energy grounds that eight colliding bubbles could form a
black hole.

It is possible to model the collision of two thin--walled bubbles by
distributional gravitational sources \cite{wu}. The bubbles start out
with flat Minkowski inside and de Sitter space outside. Each bubble is
invariant under Lorentz boosts and rotations \cite{coleman2}, therefore
the two-bubbles with a prefered axis have the symmetry O(2,1)
\cite{hawking}. This has as many killing fields as spherical symmetry
and Wu was able to demonstrate that there is a modified form of
Birkoff's theorem that can be used to find the metric.

The rapid acceleration of the bubble walls up to luminal speeds
indicates that the collision can be simplified by replacing the bubble
walls with null surfaces. When the energy from the walls is converted
into phase waves, as often happens, the energy also leaves the
collision on null surfaces. The collision of two bubbles becomes the
collision of two null shock waves, a problem whose solution is known
\cite{dt,r}.

By making these reasonable simplifications it becomes possible to
include a third or a fourth bubble in the collision. Under suitable
conditions we shall see that the energy of the bubble walls is
sufficient to refocus itself into a singularity and therefore form a
black hole or a naked singularity.

\section{O(2,1) INVARIANT METRICS}

The group O(2,1) is the symmetry group of the pseudosphere $H_2$. We
can obtain a spacetime metric which is invariant under O(2,1) by
replacing the sphere by the pseudosphere in the usual O(3) invariant
metric,
\begin{equation}
g=-S(s,x)ds^2+X(s,x)dx^2+s^2\left(d\theta^2+\sinh^2\theta
d\phi^2\right).
\end{equation}
Substituting this form of the metric into the vacuum Einstein equations
(with a cosmological constant $3/\alpha^2$), we can deduce easily that
$X=1/S=f$, where
\begin{equation}
f=1-{2M\over s}+{s^2\over\alpha^2}.
\end{equation}
The parameter $M$ is a constant but it does not have the same physical
associations of a mass. More details about these metrics can be found
in reference \cite{ehlers}.

We can draw Penrose diagrams of the surfaces that are orthogonal to the
pseudspheres. The case with vanishing cosmological constant, which will
be called pseudo--Schwartzchild, is shown in figure \ref{fig1}. This is
the usual Penrose diagram of Schwartzchild rotated through $90^0$.
Surfaces of constant coordinate $s$ are spacelike in $s>2M$ and the
metric aproaches flat space (Milne universe form) in the limit that
$s\to\infty$.

The metric with $M=0$ is de Sitter space. It will be useful to have the
embedding of the metric into a five dimensional Minkovski spacetime
with cordinates ${\bf X}=(X,Y,Z,W,V)$, where de Sitter space is defined
by the hyperboloid
\begin{equation}
{\bf X}\cdot{\bf X}=X^2+Y^2+Z^2+W^2-V^2=\alpha^2.
\end{equation}
The two de Sitter coordinate systems are related by
\begin{eqnarray}
X&=&s\sinh\theta\cos\phi,\\
Y&=&s\sinh\theta\sin\phi,\\
Z&=&(\alpha^2+s^2)^{1/2}\sin(x/\alpha),\\
W&=&(\alpha^2+s^2)^{1/2}\cos(x/\alpha),\\
V&=&s\cosh\theta.
\end{eqnarray}

During a supercooled phase transition the universe becomes locally
similar to de Sitter space. Bubbles of the true vacuum phase
(Minkowski space) appear instantaneously. Their walls move along
trajectories with O(3,1) symmetry,
\begin{equation}
({\bf X}-{\bf X}_0)\cdot({\bf X}-{\bf X}_0)=r_0^2.
\end{equation}
The radius $r_0$ is the initial bubble radius and is fixed by physical
parameters. For phase transitions at temperatures far below the Planck
temperature of $10^{19} GeV$, this radius is far smaller than the
cosmological horizon size $\alpha$.

If $r_0=0$ we can replace the bubble surface by a null surface. The
generators of this surface are light rays
\begin{equation}
{\bf X}-{\bf X}_0=\lambda\,{\bf P}\label{ray},
\end{equation}
where $\lambda$ is an affine distance along the ray and the constant
momenta ${\bf P}$ satisfy ${\bf P}\cdot{\bf P}=0$. This is the
approximation that will be used henceforth. The light rays can be also be
parameterised in an $O(2,1)$ invariant form by the substitution ${\bf
X}(s,x,\theta,\phi)$. Differentiating with respect to the affine
parameter gives
\begin{eqnarray}
\dot s&=&s^{-1}\left(ZP^Z+WP^W\right),\label{sdot}\\
\dot x&=&(sf)^{-1}\left(WP^Z-ZP^W\right),\\
\dot\phi&=&s^{-2}{\rm csh}^2\theta\left(XP^Y-YP^X\right),
\end{eqnarray}
where the components of ${\bf P}$ are denoted by $(P^X,P^Y,P^Z,P^W,P^V)$.
The values of $\dot s$ etc. are the components of the tangent vectors
in the $O(2,1)$ coordinate frame, with the latter two quantities in
brackets being the conserved linear and angular momenta respectively.

\section{TWO--BUBBLE COLLISIONS}

In this section we will take a look at the problem of two intersecting
null shells, being relevant to the collision of two bubbles when the
energy from the bubble walls leaves the collision region at the speed
of light. Numerical studies have shown that this is a good
approximation when there is a large difference of phase of the Higgs
field in each of the bubbles.

Consider a null hypersurface with null geodesic generator ${\bf l}$ and
affine parameter $u$. It is convenient to introduce a `pseudonormal'
null vector ${\bf n}=\partial/\partial v$, normalised by
${\bf l}.{\bf n}=-1$ and parallel propagated along ${\bf l}$. The
remaining tangent vectors ${\bf e_i}$, $i=1,2$, define a two dimensional
surface with induced metric
\begin{equation}
h_{ab}=g_{ab}+l_an_b+n_al_b
\end{equation}
and it is convenient to have them parallel propagated along $l$.

The embeddings of the null surface are described by extrinsic curvature
tensors $k_{ab}$ and $m_{ab}$, where
\begin{eqnarray}
k_{ab}&=&h_a^{\ c}h_b^{\ d}n_{c;d}\\
m_{ab}&=&h_a^{\ c}h_b^{\ d}l_{c;d}.
\end{eqnarray}
{}From the condition that
${\bf e_i}\cdot{\bf l}={\bf e_i}\cdot{\bf n}=0$, it is also possible to
write
\begin{eqnarray}
k_{ab}e_i^{\ a}e_j^{\ b}&=&-{\bf n}.(\nabla_{\bf i} {\bf e_j}),\\
m_{ab}e_i^{\ a}e_j^{\ b}&=&-{\bf l}.(\nabla_{\bf i} {\bf e_j})
\end{eqnarray}
Their contractions will be denoted by $k$ and $m$. It is easy to show
that $m$ is also the expansion of the null geodesic congruence, that
is $m=\nabla\cdot{\bf l}$.

A distributional source of stress--energy on the shell would take the
form
\begin{equation}
T_{ab}=\sigma l_al_b\delta(v).
\end{equation}
Junction conditions across shell can be obtained from integrating the
Einstein field equations. (See ref \cite{clarke} for example, but note
that ${\bf l}$ and ${\bf n}$ have the opposite meaning.) These force the
metric to be continuous and allow the choice of ${\bf l}$, ${\bf n}$
and ${\bf e_i}$ to be continuous. Furthermore, they relate the extrinsic
curvatures of the surfaces to the stress energy. If the surface is moving
to the right from a region $C$ into another region $B$ then,
\begin{equation}
8\pi G\sigma=k_B-k_C\hbox{,~~~and~~~}m_B=m_C.
\end{equation}
where $G$ is Newton's constant.

Two null hypersurfaces divide spacetime into four regions, labeled as
in the diagram \ref{fig2}. The null geodesic generators of the surfaces
$i=1\dots 4$ are labeled by ${\bf l_i}$ and the extrinsic curvatures will
be indexed accordingly. In the vicinity of the intersection pseudosphere,
the junction conditions imply that
\begin{eqnarray}
8\pi G\sigma_1&=&k_C^1-k^1_D\hbox{,~~}
8\pi G\sigma_2=k_C^2-k^2_B\\
8\pi G\sigma_3&=&k_B^3-k^3_A\hbox{,~~}
8\pi G\sigma_4=k_D^4-k^4_A\label{e}
\end{eqnarray}
It is also possible to obtain normal vectors from the $l$'s, for
example
\begin{equation}
{\bf n_1}=-({\bf l_1}\cdot {\bf l_2})^{-1}{\bf l_2}\hbox{~~~~~and~~~~}
{\bf n_2}=-({\bf l_1}\cdot {\bf l_2})^{-1}{\bf l_1}.
\end{equation}
These relations are specific to the regions in which the vectors are
defined, shown in fig. \ref{fig2}, but the normal vectors ${\bf n_i}$
are themselves continuous across the $i$'th surface. The continuity
equations for the $m$'s then imply the following equations for the $k$'s,
\begin{eqnarray}
({\bf l_1}\cdot{\bf l_2})k^1_C&=
&({\bf l_2}\cdot{\bf l_3})k^3_B\hbox{,~~~}
({\bf l_4}\cdot{\bf l_1})k^1_D=
({\bf l_4}\cdot{\bf l_3})k^3_A\hbox{,~~~}\\
({\bf l_3}\cdot{\bf l_2})k^2_B&=
&({\bf l_3}\cdot{\bf l_4})k^4_A\hbox{,~~~}
({\bf l_1}\cdot{\bf l_2})k^2_C=
({\bf l_1}\cdot{\bf l_4})k^4_D.
\end{eqnarray}
Finally, we allow the affine parameterisations on either side of the
collision region to differ by a constant scaling,
\begin{equation}
{\bf l_3}=\gamma\,{\bf l_1}\hbox{,~~~}
{\bf l_4}=\beta\,{\bf l_2}\label{scale}.
\end{equation}

The continuity equations can be written in a much more useful form.
Introduce
\begin{equation}
F_C=({\bf l_1}\cdot{\bf l_2})k^1_Ck^2_C,
\end{equation}
using extrinsic curvatures of surfaces adjacent to region $C$, and
similarly for each of the regions in turn. Then,
\begin{equation}
F_CF_A=({\bf l_1}\cdot{\bf l_2})
({\bf l_3}\cdot{\bf l_4})k^1_Ck^2_Ck^3_Ak^4_A,
\end{equation}
Using the continuity relations gives,
\begin{equation}
F_CF_A=({\bf l_1}\cdot{\bf l_4})
({\bf l_2}\cdot{\bf l_3})k^3_Bk^4_Dk^1_Dk^2_B=F_BF_D.
\end{equation}
This will be refered to as `the pseudo--DTR relation' \cite{dt,r}.

The metrics that are of interest to us have the O(2,1) symmetry.
Continuity of the pseudosphere sections across the null hypersurfaces
implies that the $\theta$ and $\phi$ coordinates are continuous, and
also the $s$ coordinate because it sets the area element, but the $x$
coordinate will be different on either side of the surfaces.

The null generators of the surfaces are then
\begin{equation}
{\bf l_1}={\bf e_s}-f^{-1}{\bf e_x}\hbox{,~~~}
{\bf l_2}={\bf e_s}+f^{-1}{\bf e_x}\label{null}.
\end{equation}
Using the connection components of the metric gives,
\begin{equation}
-k^1_C=k^2_C=f_C/s.
\end{equation}
Therefore the pseudo--DTR relation reads
\begin{equation}
f_Cf_A=f_Bf_D.
\end{equation}
This is the junction condition that determines the metric in any region
if the metric is known in the other three. It is also possible to get
the surface energy density from \ref{e},
\begin{equation}
\sigma_3={f_A-f_B\over 8\pi G s}\label{sigma}
\end{equation}

For the collision of two bubbles, the initial region $A$ is de Sitter
space,
\begin{equation}
f_A=1+s^2/\alpha^2.
\end{equation}
Regions $B$ and $D$ are inside the bubbles, where the metric is flat.
The energy on the bubble surface follows from equation \ref{sigma}.
Noting that the affine length along the surface is $\lambda=s$, this
can be written
\begin{equation}
\sigma=(8\pi G\alpha^2)^{-1}\lambda\label{energy}.
\end{equation}
The increase in surface energy represents the latent energy of the
false vacuum that the bubble absorbes as it expands.

The intersection region $C$ is in the true vacuum where there is no
cosmological constant and the metric is pseudo--Schwartzchild,
\begin{equation}
f_C=1-2M/s.
\end{equation}

The collision takes place at a particular value of $s$, say $s_c$. From
the pseudo--DTR relation we have
\begin{equation}
\left(1+s_c^2/\alpha^2\right)\left(1-2M/s_c\right)=1.
\end{equation}
This gives
\begin{equation}
2M={s_c^3\over\alpha^2+s_c^2}.
\end{equation}
In particular, we see that $M$ is always positive and we always get
$s_c>2M$. This is the condition for there to be {\it no} singularity
anywhere in the spacetime. Therefore, in this simplified form of the
problem, the collision of two bubbles never produces a black hole.

\section{MANY--BUBBLE COLLISION}

Although the collision of two bubbles fails to produce a black hole,
the collision can come quite close to the limit $s=2M$ where the
spacetime becomes singular. Only a small additional amount of energy
from another bubble should be required to tip the balance in favour of
a black hole. Unfortunately, introducing a third bubble breaks the
symmetry and prevents us writing down the metric explicitly. What
we can do instead is aim to find conditions under which singularities
are formed.

As a first step we shall consider the trajectories of light rays in the
spacetime formed from two colliding bubbles, especially their
refraction from the bubble walls (fig. \ref{fig3}). We shall write the
momentum vector ${\bf p}$ of a general ray in terms of the generators
of one of the bubble walls,
\begin{equation}
{\bf p}=p^u{\bf l}+p^v{\bf n}+p^1{\bf e_1}+p^2{\bf e_2}.
\end{equation}
If the light ray is itself part of a bundle from the same initial
point, then the expansion $\hat \theta$ of the bundle satisfies
Raychaudhuri's equation,
\begin{equation}
{\bf p}(\hat \theta)=-R_{ab}p^ap^b-2\hat \sigma^2-\case1/2\hat \theta^2,
\end{equation}
where $\hat\sigma$ is the shear and ${\bf p}$ has been identified with
its directional derivative following a common practice. (If we define
the analogue of $m_{ab}$ for the congruence generated by ${\bf p}$, then
$\hat\theta$ and $\hat\sigma$ are the trace and trace free parts of
$m_{ab}$.) The Einstein equations
relate the curvature to the surface energy in the bubble wall,
\begin{equation}
T_{ab}=\sigma l_al_b\delta(v),
\end{equation}
but we also have
\begin{equation}
{\bf p}(\hat \theta)=p^u{\partial\hat\theta\over\partial u}
+p^v{\partial\hat\theta\over\partial v}.
\end{equation}
Integrating gives the focusing rule,
\begin{equation}
\hat \theta_B-\hat \theta_A=8\pi G\sigma\,{\bf l}\cdot{\bf p},
\end{equation}
for a ray passing from region $A$ into region $B$.

We also see from Raychudhuri's equation that the expansion is finite,
suggesting that the momentum vector should be continuous. Nevertheless,
there is some refraction of the ray in the O(2,1) coordinate system,
due to the discontinuity of the coordinates themselves. Consider again
the ray that pases from region $A$ to region $B$, but now using the
explicit forms of the bubble generators \ref{null},
\begin{equation}
{\bf p}=\case1/2(\dot s-f_A\dot x){\bf l_3}
+\case1/2(\dot s+f_A\dot x){\bf l_4}
+\dot\theta {\bf e_\theta}+\dot\phi {\bf e_\phi}.
\end{equation}
In region $B$ we use the vector ${\bf l_2}$, related to ${\bf l_4}$ by equation
\ref{scale},
\begin{equation}
{\bf l_4}=(f_B/f_A){\bf l_2}.
\end{equation}
Restoring the $s$ and $x$ coordinates gives
\begin{equation}
\pmatrix{\dot s\cr f\dot x}_B=
\case1/2\pmatrix{1+\beta&\beta-1\cr\beta-1&1+\beta\cr}
\pmatrix{\dot s\cr f\dot x}_A,
\end{equation}
where $\beta=f_B/f_A$. Similarly, for the passage from $B$ to $C$,
\begin{equation}
\pmatrix{\dot s\cr f\dot x}_C=
\case1/2\pmatrix{1+\gamma&1-\gamma\cr1-\gamma&1+\gamma\cr}
\pmatrix{\dot s\cr f\dot x}_B,
\end{equation}
where $\gamma=f_C/f_B$.

A particular  case of interest for us later on is where the ray passes
close to the collision region, and the values of $s$ in both of the
refraction relations are the same. Then the combination gives
\begin{equation}
\pmatrix{\dot s\cr f\dot x}_C=
\pmatrix{\beta&0\cr0&\beta\cr}
\pmatrix{\dot s\cr f\dot x}_A.
\end{equation}
and
\begin{equation}
\hat \theta_C=\hat \theta_A
-8\pi G\sigma f_A^{-1}\left({\dot s}^2-f^2{\dot
x}^2\right)_A\label{fc}.
\end{equation}

Now we turn to the collision of three bubbles in a symmetrical
arrangement with centres equally spaced a proper distace $d$ appart, as
shown in figure \ref{fig4}. The cartesian coordinates can be chosen so
that the bubbles all begin at the same value the time coordinate $V$
and in the same plane. We therefore take the centres to be at,
\begin{eqnarray}
{\bf X}_1&=&(0,0,Z_1,W_1,0),\\
{\bf X}_2&=&(0,0,-Z_1,W_1,0),\\
{\bf X}_3&=&(X_3,0,0,W_3,0),
\end{eqnarray}
where
\begin{equation}
Z_1=\alpha\sin(d/2\alpha)
\end{equation}
and
\begin{equation}
X_3^2={3\alpha^2Z_1^2-4Z_1^2\over \alpha^2-Z_1^4}.
\end{equation}
We also have the condition that ${\bf X}_i\cdot{\bf X}_i=\alpha^2$ to
fix $W_1$ and $W_3$.

Bubbles $1$ and $2$ nucleate at $s=0$ in the O(2,1) coordinate system.
The bubble surfaces are given by the ray equation \ref{ray} and collide
on a surface of constant coordinate $s$,
\begin{equation}
s_c={\alpha Z_1\over\sqrt{\alpha^2-Z_1^2}}.
\end{equation}

Some elementary geometry shows that the generators of bubble $3$
intersect bubble $2$ after an affine length
\begin{equation}
\lambda_p={\alpha^2-{\bf X}_2\cdot{\bf X}_3
\over {\bf X}_2\cdot{\bf P} }.
\end{equation}
Substituting for $X_3$ gives
\begin{equation}
\lambda_p={2 Z_1^2\over W_1P^W+Z_1P^Z}.
\end{equation}
This also gives a value of the coordinate $s$ for the intersection,
\begin{equation}
s_p={\alpha Z_1\over W_1}{W_1P^W-Z_1P^Z\over W_1P^W+Z_1P^Z}.
\end{equation}

We shall restrict attention to the symmetrical rays that pass through
the collision points. For these rays, $\dot x=P^Z=0$, and from equation
\ref{sdot},
\begin{equation}
\dot s_p={P^W\over \alpha Z_1}\left(\alpha^2+Z_1^2\right).
\end{equation}
The surface energy of bubble $3$ at the intersection points is given by
equation \ref{energy}
\begin{equation}
\sigma_p={Z_1^2\over 4\pi G\alpha W_1P^W}.
\end{equation}
The expansion of the bubble wall is given by
\begin{equation}
\hat \theta_A={2\over\lambda_p}={W_1P^W\over Z_1^2}.
\end{equation}

Now it is possible to use the focussing rule \ref{fc},
\begin{equation}
\hat \theta_C=\hat \theta_A-8\pi G\sigma_p f_A^{-1} {\dot s}_p^2.
\end{equation}
This gives,
\begin{equation}
\hat \theta_C={W_1P^W\over Z_1^2\alpha^6}
\left\{\alpha^6-2Z_1^2\left(\alpha^2+Z_1^2\right)^2\right\}.
\end{equation}
Therefore the expansion of the outgoing bubble wall is negative after
the collision when $Z_1>z\alpha$, where $z\approx 0.54512$ is a root of
$z^3+z-\sqrt{1/2}=0$. If $Z_1>\alpha\sqrt{3}/2$, then the de Sitter
geometry does not allow the three bubbles to have a common collision
point and there is no refocussing.

Raychaudhuri's equation tells us that once the expansion becomes
negative it remains negative and diverges in a finite affine distance
along the ray. When the expansion diverges this forces the surface
energy to become infinite. At this point there is a spacetime
singularity, in the sense that the curvature is more singular than the
distributional curvature with which we began.

The preceeding arguments have established that refocussing occurs
along a line extending from the collision point in the direction
of the symmetrical ray. This line was chosen only for convenience
and the refraction rules obtained earlier can be used to examine
refocussing anywhere on the bubble wall.

\section{CONCLUSION}

We have argued that the bubbles nucleating at a supercooled phase
transition can be replaced by expanding null shells of energy. With
this simplification, the collision of two bubbles never produces a
singularity, in agreement with previous results.

When three bubbles collide the surface energy in parts of the walls
can be refocused to the extent that it becomes infinite. What happens
next depends on general issues of general relativity such as the cosmic
censorship conjecture. If the singularities are to be invisible
from future infinity of the emerging universe then they would have
to be hidden behind an event horizon, and form a black hole. If
cosmic censorship fails, then the bubble collision results in a
new object, but one whose time--evolution presents problems in general
relativity.

The condition for refocusing the energy along a symmetry axis
for a symmetrical collison was a proper separation $d>1.153\alpha$.
These three bubbles have a triple collision point provided that
$d<2.094\alpha$. Although the non--symmetrical case is more complicated
it is theoretically possible to follow it through using the equations
that have been given. In particular, small deviations from the
symmetrical case should leave the results unchanged.

If there are four colliding bubbles, then there is a question of
how close the boundary of the intersection region $E$ in figure
\ref{fig4} is to being future trapped. The arguments given for
three bubbles can be generalised to show that the symmetrical null
generators of this surface are converging for large bubble separations.
Of course, the surface will never be converging at the (one dimensional)
corners because of lack of smoothness. This raises two interesting
questions. If the surface where trapped, then would it be possible
to generalise the singularity theorems to conclude that the singularity
is generic, given that the strong energy condition only holds to the
future of the first collision? What is the effect of replacing a
smooth trapped surface with a piecewise smooth surface?

\acknowledgements

I am grateful to Chris Chambers for help with some of the
calculations.

\begin{figure}
\caption{Penrose diagram of the pseudo--Schwartzchild spacetime with
$M>0$. The $s$ coordinate increases from bottom to top.\label{fig1}}
\end{figure}
\begin{figure}
\caption{Two intersecting null hypersurfaces.\label{fig2}}
\end{figure}
\begin{figure}
\caption{Refraction of a light ray with momentum ${\bf p}$ by two
bubbles.\label{fig3}}
\end{figure}
\begin{figure}
\caption{The collision of three bubbles and a ray from ${\bf X_3}$ that
passes through the bubbles symmetrically.\label{fig4}}
\end{figure}
\end{document}